\documentclass[aps,superscriptaddress,preprint,nofootinbib,eqsecnum]{revtex4}
\def\half{\frac{1}{2}}

\def\be{\begin{equation}}
\def\ee{\end{equation}}
\def\bea{\begin{eqnarray}}
\def\eea{\end{eqnarray}}

\def\nott#1{\setbox0=\hbox{$#1$}                % set a box for #1
   \dimen0=\wd0                                 % and get its size
   \setbox1=\hbox{/} \dimen1=\wd1               % get size of /
   \ifdim\dimen0>\dimen1                        % #1 is bigger
      \rlap{\hbox to \dimen0{\hfil/\hfil}}      % so center / in box
      #1                                        % and print #1
   \else                                        % / is bigger
      \rlap{\hbox to \dimen1{\hfil$#1$\hfil}}   % so center #1
      /                                         % and print /
   \fi}                                         %
\usepackage{epsf} 
\usepackage{graphicx,amsmath,amssymb,bm}

\begin{document}

\author{Kiyoumars A. Sohrabi}
\email{sohrabi@itp.unibe.ch}
\affiliation{Albert Einstein Center for Fundamental Physics, University of Bern,
Sidlerstrasse 5, 3012 Bern, Switzerland}

\begin{abstract}
It is usually believed that unlike the external magnetic field which one can set 
directly, vorticity is a property of the flow of particles, which is indirectly controlled
by external fields and initial conditions. Using the curved-space technics it is shown that
the influence of the vorticity on the relativistic chiral fermions can indeed be controlled directly.  

\end{abstract}

\title{Microscopic Study of Vorticities in Relativistic Chiral Fermions}

\maketitle
%%%%%%%%%%%%%%%%%%%%%%%%%%%%%%%%%%%%%%%%%%%%%%%%%%%%%%%%%%%%%%%%%%%%%%%%%%%%%%%%%%%%%%%%%%%%%%%%%
%%%%%%%%%%%%%%%%%%%%%%%%%%%%%%%%%%%%%%%%%%%%%%%%%%%%%%%%%%%%%%%%%%%%%%%%%%%%%%%%%%%%%%%%%%%%%%%%
%%%%%%%%%%%%%%%%%%%%%%%%%%%%%%%%%%%%%%%%%%%%%%%%%%%%%%%%%%%%%%%%%%%%%%%%%%%%%%%%%%%%%%%%%%%%%%%%%%
%%%%%%%%%%%%%%%%%%%%%%%%%%%%%%%%%%%%%%%%%%%%%%%%%%%%%%%%%%%%%%%%%%%%%%%%%%%%%%%%%%%%%%%%%%%%%%%%%%%
\section{Introduction}
  Field theories under the influence of external sources  have been studied since a long time ago. One important
application of the quantization of these theories goes back to Schwinger in 1951 \cite{Schwinger}.
He considered the effect of  background electromagnetic fields on \text{$\gamma$}-decay of
neutral mesons. His technics were developed further and  got various
applications in relativistic hadron-hadron collisions \cite{Bleicher}, 
QCD phenomenology using chiral effective Lagrangians \cite{Hatsuda}
and gravitational anomalies \cite{Witten}. As such is  chiral magnetic effect (CME), introduced by Kharzeev \emph{et al.} 
\cite{Kharzeev}. This effect occurs when a nonzero current is produced  along the magnetic
field in a chiral medium.
Separate computations to approximate its size has been done \cite{Kharzeev-Son} 
and comparison with observables such as charge-dependent correlations \cite{Abelev} 
provided a good foundation for further analysis.

A closely related exciting problem to the CME is chiral vortical effect (CVE) where a current appears along 
the axis of vorticity. Historically, it was first realized by Vilenkin in 1980 \cite{Vilenkin:1980ft}, 
while  complete understandings of the hydrodynamical current 
and its connection to the triangle anomaly was given by Son and Surowka \cite{Surowka}.
It has been studied in the context of gauge/gravity duality \cite{Erdmenger}, non-zero 
chemical potential \cite{Zhitnitsky}, group theory \cite{Roy}, kinetic theory \cite{Chen:2014cla} and
topological invariants in momentum space \cite{Volovik}.
Its applications are numerous; in principle all phenomena that include
magnetic fields, influence of vorticities can also be implemented into the medium. 
Future application of the CVE will shed light  on  3D graphene  and Weyl semimetals \cite{Turner}. Further
relevant topics in this context are superfluids and Fermi liquids \cite{Iida, Kirilin, Xiong}.

Although a lot has been done in the literature to understand CME in the recent years, but the same treatment
is lacking in the context of CVE. For a few computations for instance refer to 
\cite{Kalaydzhyan:2014bfa}.  In fact, there are many contradictions about these anomaly-induced effects
and their physical picture \cite{Landsteiner}. For a review on related issues refer to \cite{Loganayagam}. 
We show using standard methods that CVE
is comprehensible in the context of quantum field theory and deducible from first principles
self-consistently.

First, let's mention an issue with \emph{causality}; the common definition of the vorticity from
classical mechanics is $\bm \omega=\half\nabla\times \bm u$.
Consider
fluid's motion confined to a plane in two
dimensions. Then 
$\omega_{z}=\frac{1}{r}\frac{\partial}{\partial r}(ru_{\phi})$,
where $(r,\phi)$ are 2D cylindrical coordinates.
We are looking for a setup that the vorticity
is zero at the origin at some initial time thus we
choose $u_{\phi}=a r^{n}$. Then
$\omega_{z}=a(n+1)r^{n-1}$ for $n>1$. That is,
the vorticity gets larger as we go further away from the
origin and eventually fluid's speed becomes larger
than the speed of light.

One way to overcome this problem is to promote the Minkowski's flat spacetime to a 
curved spacetime. To see this, note that the Poincar\'{e}
group leaves the Minkowski metric in flat space
invariant. Infinitesimal transformation of this 
group has the form, $y^{'a}=y^{a}+\xi^{a}\epsilon$.
Invariance of the Minkowski metric then means that 
$\partial_{c}y^{'a}\partial_{d}y^{'b}\eta^{cd}=\eta^{ab}$.
The key point is that an arbitrary boost of the form $x^{\mu}=x^{\mu}(y^{'})$
satisfies $\xi^{\mu}(x)=\partial_{a^{'}}x^{\mu}\xi^{a^{'}}(y^{'})$.
Under this transformation the Minkowski metric is no
longer invariant 
$g^{\mu\nu}=\partial_{a^{'}}x^{\mu}\partial_{b^{'}}x^{\nu}\eta^{ab}$.

In recent years, applications of curved spacetime backgrounds as a
probe for symmetries of the underlying flat-spactime field theories have become popular.
The basic idea is very simple, intuitively whenever we want to study the effect of 
electromagnetism, we add gauge potentials
to  the Lagrangians of the field theories. Similarly, the study
of a system under relativistic rotations or vorticities
is possible by a specific choice of a background metric.
In what follows, we consider
perturbative background perturbations of
the type $g_{\mu\nu}(x)=\eta_{\mu\nu}+h_{\mu\nu}(x)$. 
Generally, we can classify these attempts in two groups of macroscopic and microscopic physics. 

%%%%%%%%%%%%%%%%%%%%%%%%%%%%%%%%%%%%
%%%%%%%%%%%%%%%%%%%%%%%%%%%%%%%%%%%%
%%%%%%%%%%%%%%%%%%%%%%%%%%%%%%%%%%%%
\emph{The macroscopic physics.}\textemdash 
Perturbative technics with the background metric have already been used for deriving Kubo
formulas  in the relativistic hydrodynamics \cite{Baier}. As an example,
it has been shown that in a conformal parity-even theory, the hydrodynamic transport 
coefficient that couples vorticities in the stress tensor  is given by
\begin{eqnarray}
  \label{viscosity}
T^{\mu\nu}  = T_{\text{ideal}}^{\mu\nu}
-\lambda_{3}\Omega_{\lambda}{}^{\langle\mu}\Omega^{\nu\rangle\lambda}
  \,, 
\end{eqnarray}
with $\Omega^{\mu\nu}  \equiv  \half \Delta^{\mu\alpha} \Delta^{\nu\beta}
( \nabla_\alpha u_\beta - \nabla_\beta u_\alpha)$, if for simplicity we assume
the shear tensor, 
$\sigma^{\alpha\beta}\equiv  2 \nabla^{\langle \alpha } u^{\beta \rangle}=0 $.
Here we are using the convenient notation, $A^{\langle \mu \nu \rangle} \equiv
\frac{1}{2} \Delta^{\mu\alpha} \Delta^{\nu\beta}
( A_{\alpha\beta} + A_{\beta\alpha} )
- \frac{1}{3} \Delta^{\mu\nu} \Delta^{\alpha\beta} A_{\alpha\beta}$, that 
subtracts the trace of the tensor if we choose $\Delta^{\mu\nu}=g^{\mu\nu}+u^{\mu}u^{\nu}$.
The above information is sufficient to determine the corresponding Kubo formula \cite{Moore,Arnold}, 
\be
\label{lambda3}
\lambda_3 = -4 \lim_{p^z,q^z \rightarrow 0}
\partial_{p^z} \partial_{q^z}
G^{xy,xt,yt}_{E}(p,q)\,,
\ee
where here $G^{xy,xt,yt}_{E}$ is defined in \cite{Sohrabi} as
\begin{eqnarray} \label{action_E} G_{\rm E}^{xy,xt,yt}(p,q)
 &\equiv& \left. \int d^4 xd^4 y
    e^{-i(p\cdot x+q\cdot y)}\right.{}
\nonumber\\
  & & {}\left. \times
\frac{2^3 \; \partial^3 \ln {\rm Z}}
  {\partial g_{yt}(x)  \partial g_{xt}(y) \partial g_{xy}(0)}
\right|_{g_{\mu\nu} = \delta_{\mu\nu}}{}\,,
\end{eqnarray}
 with
\begin{eqnarray}
Z\left[g_{\mu\nu}\right]=\int\mathcal{D}\phi  \; \exp
\left\{-S_{\rm E}\left[\phi,g_{\mu\nu}\right]\right\} \,,
\end{eqnarray}
which basically what it's saying is that, $\lambda_{3}$ is given by the second derivatives of the 
three-point correlation function of the stress tensors plus some contact terms.
For a derivation of this transport coefficient directly from
partition function refer to \cite{Minwalla} or \cite{Kovtun}. The lesson to take home
from the above example is that specific choice of the metric components, 
in this case $h_{0x}(z)$ and $h_{0y}(z)$, gives nonzero values for $\Omega_{xz}=−\half \partial_{z}h_{ox}$
and $\Omega_{yz}=−\half \partial_{z}h_{oy}$.

In parity-odd theories, on the other hand, vorticity as an axial vector
appears at first order in the derivative expansion in the current $j^{\mu}$
 \cite{Surowka},
\begin{equation}
    j^{\mu} = nu^{\mu}-\sigma T\left(g^{\mu\nu}
    +u^{\mu}u^{\nu}\right)\partial_{\nu}\left(\frac{\mu}{T}\right)
    +\xi\omega^{\mu}\,,
    \quad 
    \omega^{\mu} =
    \epsilon^{\mu\nu\lambda\rho}u_{\nu}\Omega_{\lambda\rho}\,,
\end{equation}
and the Kubo formula for the anomalous coefficient $\xi$, reads as,  
 \be
\xi =-i \lim_{p^y\rightarrow 0}
\partial_{p^y}
G^{ty,z}_{E}(p)\,,
\ee
where the above Euclidean Green's function is expressed in terms of
the second derivatives of the partition function. The first derivative is with respect to
 $A^{z}$, the $z$ component of the electromagnetic four-potential,  then the second derivative with 
respect to $g^{ty}$. This works since as one can check, the lower-indexed
 rest frame  will be $u_{\mu}=(-1,h_{ox},0,0)$. Plugging this value
into the definition of the vorticity, we obtain $\omega^{z}=-\half\partial_{y}h_{0x}$.

%%%%%%%%%%%%%%%%%%%%%%%%%%%%%%%%%%%%
%%%%%%%%%%%%%%%%%%%%%%%%%%%%%%%%%%%%
%%%%%%%%%%%%%%%%%%%%%%%%%%%%%%%%%%%%
\emph{The microscopic  physics.}\textemdash With a self-consistent approach to study
the relativistic vorticity, we will be capable of uncovering new properties of 
field theories in various mediums.
The most significant examples are fermionic systems in which calculations
are naturally more involved. For simplicity, we concentrate
on massless chiral fermions. As will be exhibited,  alterations
to a medium with the vorticity are twofold; It's required to substitute  
flat-spacetime derivatives with covariant derivatives that include  couplings of the
vorticity to the spin. Technically this term is called spin connection.
Furthermore, spacial components of the Dirac gamma matrices get boosted along 
the metric as $\gamma^{i}-\beta^{i}\gamma^{0}$, where  we can interpret
$\beta^{i}$ as the speed of relative local patches that cover the spacetime.
The prescribed modifications are dictated by the covariance of the field theory
in curved spacetime and they will serve a significant role in our discussion. As we shall see,
momenta of the spinors are subject to a shift by $\pm\lambda\omega^{z}$ 
with $\lambda=\half$, depending on whether the spinors are right-handed or left-handed and 
also the direction of the vorticity. As we will observe, the interpretation employs a
coupling between the orbital angular momentum and the spin. In virtue of the covariance of the Dirac
equation, dispersion relations acquire a new kinetic term of the form $\beta^{i}p_{i}$, with $p_{i}$ the
momenta of the spinors.

Special focus is on chiral fermions. Our goal is to systematically expand the background perturbations
 to first order  and derive the equations of motion and the corresponding solutions of the
Dirac equation. Along the way, we study the Hamiltonian and the chiral current 
from the first principles. This is done in the most convenient way  through the second quantization. 
This will allow us to  comprehend the separation of
the right-handed or left-handed fermions under the influence of an external vorticity microscopically.
To embody the later, primarily we study the effects of the chemical potential and the background
electromagnetic potentials on  interactions that are present in the theory.

In Section~\ref{sec1}\,. we give a short but complete review of the formalism for studying
fermions in an arbitrary curved background. Starting from the Lagrangian, we derive the equations of motion
and the Hamiltonian. Section~\ref{sec2}\,. will be the applications of the former tools. 
In \ref{subsec2}\,. we shall square the Hamiltonian operator and investigate 
the interactions that appear and hence, we infer properties of the eigenvectors and
eigenvalues of the Dirac equation. In \ref{subsec3}\,. we look into the modifications 
of the boost and rotation operators produced by vorticities. Finally in \ref{subsec4}\,. 
we compute the chiral vortical current. The summary will recap our main ideas
and results.

%%%%%%%%%%%%%%%%%%%%%%%%%%%%%%%%%%%%%%%%%%%
%%%%%%%%%%%%%%%%%%%%%%%%%%%%%%%%%%%%%%%%%%%%
%%%%%%%%%%%%%%%%%%%%%%%%%%%%%%%%%%%%%%%%%%%
\section{Review of the Tools}\label{sec1}
Our goal is to find the Hamiltonian and the equations
of motion for free massless fermions in a medium with  vorticities
with the aid of curved-background methods. To this end, we start with a general
formalism and then later on we concentrate on the specific background perturbations
that produce vorticities. In this section, we present a brief and self-contained
review of the prerequisite material  based on \cite{Dewitt}.
\subsection{Local Lorentz Frames }
It is convenient to describe fermions
in a curved space using local Lorentz frames
$\bm e_{\alpha}$, on a space-time
manifold $\mathcal{M}$ which meets the condition  
$\bm e_{\alpha}\cdot \bm e_{\beta}=\eta_{\alpha\beta}$
, with $\eta_{\alpha\beta}$ being the 4D
Minkowski metric with the signature of $-+++$. 
The inner product on the above manifold
is defined with respect to a metric $g_{\mu\nu}$. There,
 written in components has the form  
$e_{\alpha\mu}e^{\,\mu}_{\beta}=\eta_{\alpha\beta}$ that also satisfies
$e_{\alpha\mu}e^{\alpha}_{\,\nu}=g_{\mu\nu}$.
Throughout this section we stick to the following  notation for the indices; coordinate indices are chosen
from the middle of the Greek alphabet and are lowered and raised by the metric $g_{\mu\nu}$ and
its inverse $g^{\mu\nu}$, supplemented by the indices from the first of the alphabet which
are called frame indices and are handled by the Minkowski metric $\eta_{\alpha\beta}$ and its inverse 
$\eta^{\alpha\beta}$.

In order to find the covariant derivatives for spinors we need to consider both of
the general coordinate and the Lorentz transformations.
In the curved background, derivative of a contravariant vector field $\Psi^{\mu}$ takes the from
$\Psi^{\mu}_{;\nu}=\Psi^{\mu}_{,\nu}+\Gamma^{\mu}_{\ \ \sigma\nu}\Psi^{\sigma}$,
with $\Gamma^{\mu}_{\,\sigma\nu}$ the Riemann connection 
$\Gamma_{\sigma\nu\mu}=\half(g_{\sigma\nu,\mu}+g_{\sigma\mu,\nu}-g_{\nu\mu,\sigma})$.
On the other hand, the same vector field in the local Lorentz frame is given by the contraction of
$\Psi^{\alpha}=e^{\alpha}_{\ \mu}\Psi^{\mu}$.
Then, its covariant derivative can equally well be written in this frame as
$\Psi^{\alpha}_{\ ;\mu}=\Psi^{\alpha}_{\ \,\mu}+\Sigma^{\alpha}_{\ \beta\mu}\Psi^{\beta}$,
where $\Sigma^{\alpha\beta}_{\mu}$ is known as the spin connection. Thus, relating these two
derivatives requires the covariant differentiation to act neutral in passing the coordinate
indices  to the local Lorentz frame indices,
\begin{equation}\label{DummyFrame}
    \Psi^{\alpha}_{\ ;\mu}=
    e^{\alpha}_{\ \nu}\Psi^{\nu}_{\ ;\mu}
    =e^{\alpha}_{\ \nu}\left(\Psi^{\alpha}_{\ ,\mu}+\Gamma^{\nu}_{\ \ \sigma\mu}\Psi^{\sigma}\right)\,,
\end{equation}
comparison of the two derivatives leads us to
\begin{equation}
    \Sigma^{\alpha\beta}_{\mu}=
    e^{\alpha\sigma}e^{\beta\nu}\Gamma_{\sigma\nu\mu}-e^{\alpha}_{\ \nu,\mu}e^{\beta\nu}\,.
\end{equation}

It's also easy to show that $\Sigma^{\alpha\beta}_{\mu}$ is antisymmetric
when  $\alpha$ and $\beta$ are interchanged.
One usually uses the Dirac matrices that satisfy 
$\left\{\gamma_{\alpha},\gamma_{\beta}\right\}=2\eta_{\alpha\beta}$
in the local Lorentz frame where its contravariant vector takes the form
$\gamma^{\mu}=\gamma_{\alpha}e^{\alpha\mu}$, with $\left\{\gamma^{\mu},\gamma^{\nu}\right\}=2g^{\mu\nu}$.
Thereby, the covariant derivative of the fermionic fields in terms of the spin connection reads 
\be\label{nabla}
    \nabla_{\mu}\Psi = \partial_{\mu}\Psi+\half G_{[\alpha\beta]} \, \Sigma^{\alpha\beta}_{\,\,\mu} \, \Psi
    \,,\quad
    \nabla_{\mu}\bar{\Psi} = \partial_{\mu}\bar{\Psi}-\half\bar{\Psi} \, G_{[\alpha\beta]} \, 
    \Sigma^{\alpha\beta}_{\,\,\mu}\,,
\ee
with $G_{[\alpha\beta]}=\frac{1}{4}[\gamma_{\alpha},\gamma_{\beta}]$, 
$\bar{\Psi}=\Psi^{\dagger}\eta$ and $\eta=i\gamma^{0}$.
%%%%%%%%%%%%%%%%%%%%%%%%%%%%%%%%%%%%%%%%%%%%%%%%%
%%%%%%%%%%%%%%%%%%%%%%%%%%%%%%%%%%%%%%%%%%%%%%%%%
%%%%%%%%%%%%%%%%%%%%%%%%%%%%%%%%%%%%%%%%%%%%%%%%%%
\subsection{The Equations of Motion}
Now that we have the form of the covariant derivative
of the spinors, we can write the density of the Lagrangian
for the massless spinors as 
\begin{equation}
\label{Dirac-log}
    \mathcal{L}=-ig^{1/2}\bar{\Psi}\gamma^{\mu}
    \left(\frac{\partial}{\partial x^{\mu}}+\half G_{[\alpha\beta]}\Sigma^{\alpha\beta}_{\quad\mu}\right)\Psi\,.
\end{equation}
For future advantageous, we will rewrite the above
Lagrangian density using the Leibniz's rule,
\be\label{leib}
    g^{1/2}\nabla_{\mu}\bar{\Psi}\gamma^{\mu}\Psi+g^{1/2}\bar{\Psi}\gamma^{\mu}\nabla_{\mu}\Psi
    =\nabla_{\mu}\left(g^{1/2}\bar{\Psi}\gamma^{\mu}\Psi\right)\,,
\ee
that follows from the simple identity
$[\gamma_{\alpha},G_{[\beta\gamma]}]\;\Sigma^{\beta\gamma}_{\quad\mu}
=2\Sigma_{\alpha\;\;\mu}^{\;\gamma}\;\gamma_{\gamma}$. In view of the
right-hand side of Eq.\,(\ref{leib}) that enjoys 
$\nabla_{\mu}\left(g^{1/2}\bar{\Psi}\gamma^{\mu}\Psi\right)=
\partial_{\mu}\left(g^{1/2}\bar{\Psi}\gamma^{\mu}\Psi\right)$, 
we can compute both sides separately to arrive at
\be\label{F-identity}
    \bar{\Psi}\partial_{\mu}\gamma^{\mu}\Psi
    =\half\bar{\Psi}[\gamma^{\mu},G_{[\alpha\beta]}]\Sigma^{\alpha\beta}_{\quad\mu}\Psi\,.
\ee
The Dirac Lagrangian in Eq.\,(\ref{Dirac}) now, takes the form 
\be
    \mathcal{L}=-ig^{1/2}\;\bar{\Psi}
    \left[\half\left\{\gamma^{\mu},\frac{\partial}{\partial x^{\mu}}\right\}
    +\frac{1}{4} \left\{\gamma^{\mu},G_{[\alpha\beta]}\right\}\Sigma^{\alpha\beta}_{\quad\mu}\right]\Psi\,,
\ee
and similarly the equations of motion in curved background will be
\begin{equation}\label{Dirac}
    \left[\half\left\{\gamma^{\mu},\frac{\partial}{\partial x^{\mu}}\right\}
    +\frac{1}{4}\left\{\gamma^{\mu},G_{[\alpha,\beta]}\right\}\Sigma^{\alpha\beta}_{\quad\mu}\right]\Psi=0\,.
\end{equation}
At this point, it's simpler to decompose the metric in the following way  
\begin{equation}\label{metric-matrix}
    g_{\mu\nu}=\left( 
    \begin{array}{cc}
    -\alpha^2+\beta_{k}\beta^{k} & \beta_{j}  \\
    \beta_{i} & \eta_{ij}  \\
    \end{array} \right)\,,
\quad 
    g^{\mu\nu}=\left( 
    \begin{array}{cc}
    -\alpha^{-2} & \alpha^{-2}\beta^{j}  \\
    \alpha^{-2}\beta^{i} &\ \eta_{ij}-\alpha^{-2}\beta^{i}\beta^{j}  \\
    \end{array} \right)\,,
\end{equation}
where in the above matrices we used  $\beta^{i}=\eta^{ij}\beta_{j}$, $g=\alpha^{2}$
and  $i,j, k\in\{1,2,3\}$. Here $\alpha$ is a positive function of the coordinates $x^{i}$ 
known as the lapse function and $\beta^{i}$ is called shift vector. 

When spacetime is flat, the easiest choice for the components of the  local
Lorentz frames is  simply given by $e_{\alpha\mu}=\eta_{\alpha\mu}$. To adopt a decomposition
in curved space, we specialize our study to  \emph{stationary} backgrounds. But prior to that,
let's indicate what we mean by the stationary backgrounds. When the metric $g_{\mu\nu}$
and thus the dynamics of the spinors are independent of time together with the condition
$|\beta^{i}|<1$, the Killing vector associated with the time translation will be timelike.
This is the main assumption of this paper.
 
The fact that the time direction is the Killing vector makes it possible to define local
orthonormal frames $\bm e_{a}$, $a=1,2,3$ on each patch of the 3D spatial section 
$x^{0}=t=\text{constant}$, that is  $e_{a}^{\ i}e_{bi}=\eta_{ab}$ and 
$e_{ai}e^{a}_{\ j}=\eta_{ij}$, where  $a, b\in\{1,2,3\}$.

One then drags each field $\mathbf{e}_{a}$ by
the vector $\partial/\partial x^{0}$, to
extend the previous orthonormal construction
to 4D. All we need to add is $e_{a}^{\ 0}=0$
and $e_{a0}=\beta_{i}e_{a}^{\ i}$ 
and an extra vector field $\mathbf{e}_{0}$ with components
$e_{0}^{\ 0}=\alpha^{-1}$ and 
$e_{0}^{\ i}=-\alpha^{-1}\beta^{i}$ 
in addition to $e_{00}=-\alpha$ and $e_{0i}=0$.
We summarize this foliation of spacetime  in
TABLE\,\ref{Lorentz-frame}.

\begin{table}
    \centering
    \begin{tabular}{ |c| c| c| }
\hline
    Vierbein's components  & Time-like coordinate dir.       & Space-like coordinate dir.             \\
\hline   
    Time-like Lorentz dir. & $e_{0}^{\;\; 0}=\alpha^{-1}$     & $e_{0}^{\;\; i}=-\alpha^{-1}\beta^{i}$ \\
                          & $e_{00}=-\alpha$                 & $e_{0i}=0$                             \\
\hline
    Space-like Lorentz dir. & $e_{a}^{\;\; 0}=0$               & $e_{a}^{\;\; i}e_{bi}=\eta_{ab}$       \\
                          & $e_{a0}=\beta_{i}e_{a}^{\;\; i}$ & $e_{ai}e^{a}_{\;\; j}=\eta_{ij}$          \\
\hline
    \end{tabular}
    \caption{Decomposition of local Lorentz frames}
    \label{Lorentz-frame}
\end{table}

The Hamiltonian is calculated using the usual definition of $H$
as the Legendre transformation of $\mathcal{L}$. This operation yields 
\be\label{hamilton-exwi}
    H=\int d^3\bm x\,\alpha \, \bar{\Psi}
      \left[
      \half
	    \left\{
	    \gamma^{i}-\alpha^{-1}\beta^{i}\gamma^{0},\frac{\partial}{\partial x^{i}}
	    \right\}
      +\frac{1}{4}
      \left\{\gamma^{\mu},G_{[\alpha\beta]}\right\}\Sigma^{\alpha\beta}_{\quad\mu}
      \right]
      \Psi\,,
\ee
here $\gamma^{\mu}=\gamma^{\alpha}e_{\alpha}^{\;\;\mu}$ and $\gamma^{0}$ has the
frame index 
\footnote{We use 
\bea
\gamma^{0}=\left( 
    \begin{array}{cc}
    0 & -i  \\
    -i & 0  \\
    \end{array} \right)\,,
    \quad
\gamma^{a}=\left( 
    \begin{array}{cc}
    0 & -i\sigma^{a}  \\
    i\sigma^{a} & 0  \\
    \end{array} \right) \quad \text{with}\quad
\gamma^{5}=\left( 
    \begin{array}{cc}
    I & 0  \\
    0 & -I  \\
    \end{array} \right)\,.
\eea    
}. To write the first therm in the above bracket, we used TABLE\,\ref{Lorentz-frame}. 
It worth pointing out that this Hamiltonian is exact. In the next section, we shall narrow
our study and only consider specific choices of the metric that meets $\Sigma^{xy}_{0}\neq0$. 
In addition, we will be interested to know the effect of chemical potentials.
%%%%%%%%%%%%%%%%%%%%%%%%%%%%%%%%%%%%%%%%%
%%%%%%%%%%%%%%%%%%%%%%%%%%%%%%%%%%%%%%%%%%
%%%%%%%%%%%%%%%%%%%%%%%%%%%%%%%%%%%%%%%%%%
\section{Applications}\label{sec2}
\subsection{Vortical Dipole Moment}\label{subsec1}
In the previous section, we mentioned the exact
Hamiltonian for massless free spinors in
curved space. Since we are interested in first order
perturbations of the background metric, we  simplify
Eq.\,(\ref{hamilton-exwi}) with
the metric given in Eq.\,(\ref{metric-matrix}) and assume $\partial_{i}\beta^{i}=0$ to deduce
\be\label{hamilto-pert}
    H=\int d^3\bm x\,\bar{\Psi}
      \left[
	    \left(-\alpha^{-1}\beta^{i}\gamma^{0}+\gamma^{i}\right)\frac{\partial}{\partial x^{i}}
	     +\frac{1}{4} \left\{\gamma^{\mu},G_{[\alpha\beta]}\right\}
	     \Sigma^{\alpha\beta}_{\quad\mu}+i\gamma^{0}\mu
      \right]
      \Psi+\mathcal{O}(\alpha^2,\beta^2)\,,
\ee

Let's pause for a moment and compare
the Hamiltonian in Eq.\,(\ref{hamilto-pert}) with what we already know
from the flat spacetime. As one can notice,  adding chemical potential $\mu$ can be confusing; it
couples to the conserved charge and the fact that whether we need
to write it in terms of $\gamma^{0}$ or $\gamma_{0}$ depends
on the normal vector to the equal-time surfaces. We come back to this point and explain more 
when we want to solve the Dirac equation. In view of the flat space, 
$\mathcal{H}^2$ (the operator form of $H^2$) has less information and  simpler
form than the Dirac Hamiltonianin by comparison. We are curious to see a similar expression
for $\mathcal{H}^2$ in curved space.
In the former background metric we have (with $\alpha=1$),
\be\label{H-first-gen}
    \mathcal{H}^2=
    \eta^{-1}\left(
    \gamma^{i}\frac{\partial}{\partial x^{i}}
    +\frac{\gamma^{\mu}}{2} G_{[\alpha\beta]}\Sigma^{\alpha\beta}_{\quad \mu}
    +i\mu\gamma^{0}
    \right)\eta
    \left(
    \gamma^{j}\frac{\partial}{\partial x_{j}}
    +\frac{\gamma^{\nu}}{2} G_{[\alpha\beta]}\Sigma^{\alpha\beta}_{\quad\nu}
    +i\mu\gamma^{0}
    \right)\,,
\ee
where here none of the gamma matrices have the Lorentz index (i.e. frame index). The expansion of the product 
gives
\bea
    \mathcal{H}^2&=&\eta^{-1}\gamma^{i}\eta\gamma^{j}\nabla_{i}\nabla_{j}
		  +\half\eta^{-1}\gamma^{i}\eta\gamma^{0}G_{[\alpha\beta]}\Sigma^{\alpha\beta}_{0}\nabla_{i}
		  +\half\eta^{-1}\gamma^{0}G_{[\alpha\beta]}\eta\Sigma^{\alpha\beta}_{0}\gamma^{j}\nabla_{j}
		  +i \, \mu \, \eta^{-1}\gamma^{i}\eta\gamma^{0}\nabla_{i}
\nonumber\\&&
		  +i \, \mu \, \eta^{-1}\gamma^{0}\eta\gamma^{j}\nabla_{j}
		  +\frac{i}{2} \, \mu \, \eta^{-1}\gamma^{0}\eta\gamma^{0} \, G_{[\alpha\beta]}\Sigma^{\alpha\beta}_{0}
		  +\frac{i}{2} \, \mu \, \eta^{-1}\gamma^{0}G_{[\alpha\beta]}\eta\gamma^{0}\Sigma^{\alpha\beta}_{0}
		  -\mu^2\,,
\eea
in arriving at the later form, we have separated the spacial components of 
$\gamma^{\mu}$ in the second terms of the parentheses in Eq.\,(\ref{H-first-gen}) 
and combined them with partial derivatives to construct the spacial components of
the covariant derivatives.
As we mentioned in the introduction, consider a situation with the constant vorticities
parallel to the $z$-axis, then one configuration would be provided by $\beta^{x}(y),\beta^{y}(x)\neq0$.
 One can check that the non-zero components of the spin connection are then given by
\bea\label{setup}
    \Sigma^{0y}_{\quad x}&=&-\frac{1}{2}\big(\partial_{x}\beta^{y}+\partial_{y}\beta^{x}\big)\,,
\nonumber\\
    \Sigma^{0x}_{\quad y}&=&-\frac{1}{2}\big(\partial_{x}\beta^{y}+\partial_{y}\beta^{x}\big)\,,
\nonumber\\
    \Sigma^{xy}_{\quad 0}&=&\;\half\;\big(\partial_{y}\beta^{x}-\partial_{x}\beta^{y}\big)\,.
\eea
If we define $\omega^{z}\!\equiv\!-\half\partial_{y}\beta_{x}$ and assume a symmetric setup
that has the property of $-\partial_{x}\beta^{y}\!=\!\partial_{y}\beta^{x}$, then the above
expressions reduce to $\Sigma^{0y}_{\quad x}\!=\!0$, $\Sigma^{0x}_{\quad y}\!=\!0$ and 
$\Sigma^{xy}_{\quad 0}\!=\!-2\omega_{z}$. 
The prefactor of two in the later result can be interpreted as having
double vorticities in the same direction.

With the utility of $\nabla_{i}\nabla_{j}-\nabla_{j}\nabla_{i}
=-\half G_{[\alpha\beta]}\,R^{\alpha\beta}_{\quad ij}$, we shall symmetrize
the curved space gradients in Eq.\,(\ref{hamilto-pert}) for 
$R^{0j}_{\quad kl}=\partial_{j}\partial_{[k}\beta_{l]}$ vanishes in the limit 
of constant vorticities. After carrying out  necessary algebraic rearrangements, we infer that
\bea\label{main-Hamilton}
    H^2&=&\int d^3\bm x\,
    \bar{\Psi}\Big[
    \left(P_{x}-2i\omega_{z}\gamma^{y}\gamma^{0}\right)^2
    \!+\!
    \left(P_{y}+2i\omega_{z}\gamma^{x}\gamma^{0}\right)^2
    \!+\!
    P_{z}^2 
\nonumber\\&&\hspace{4cm}
    +4\left(\beta^{x}P_{x}+\beta^{y}P_{y}-\mu\right) G^{[0a]}P_{a}-2\mu\,\omega_{z}\sigma^{z}-\mu^2\Big]\Psi
    \,,
\eea
where $G^{[0a]}=\frac{1}{4}\left[\gamma^{0},\gamma^{a}\right]$
 and $P_{i}=-i\partial_{i}$. As we will discuss more carefully in the next section, the energy eigenvalues 
are shifted by $\beta^{i}P_{i}$, therefore naturally cross terms with $\beta^{i}$
appear in the eigenvalues of $H^2$.   The  term proportional to $\sigma^{z}$ in 
Eq.\,(\ref{main-Hamilton}) is understood better in the Weyl representation
\be
-\mu~2\omega^{z}\left(\Psi^{\dagger}_{R}\Psi_{L}-\Psi^{\dagger}_{L}\Psi_{R}\right)\,.
\ee
This term, $\mu~\bm\omega\cdot\bm\sigma$, is identical to the magnetic 
dipole moment in the  $\bm\mu_{B}\cdot\bm B$ with $\bm \mu_{B}=\frac{\bm\sigma}{2}$ for spin-$\half$ fermions
in an external background
magnetic field. In this case \emph{vortical dipole moment} is defined by $\mu_{V}=\mu\,\bm\sigma$.

For the case that $\mu=0$, we can understand the shifts in the momenta $P_{x}$ and $P_{y}$. 
The $z$ component of the total angular momentum, $\bm J=\bm L+\bm S$, 
commutes with $H^{2}$ whence $J_{z}$ is conserved and its eigenvalues, $m$,
can be used for labeling the states of the system i.e. $J_{z}|\Psi\rangle=m|\Psi\rangle$.
 
Recalling the configuration in Eq,\,(\ref{setup}), we can think of a setup that constitutes
two vorticities parallel but in the opposite directions. This is achieved by the condition 
$\partial_{x}\beta^{y}=\partial_{y}\beta^{x}$. Thus, the squared of the Hamiltonian will take the form
\bea
    H^2&=&\int d^3\bm x\,
    \bar{\Psi}\Big[
    \left(P_{x}-2i\omega_{z}\gamma^{y}\gamma^{0}\right)^2
    \!+\!
    \left(P_{y}-2i\omega_{z}\gamma^{x}\gamma^{0}\right)^2
    \!+\!
    P_{z}^2 
\nonumber\\&&\hspace{4cm}
    +4\left(\beta^{x}P_{x}+\beta^{y}P_{y}-\mu\right) G^{[0a]}P_{a}-\mu^2\Big]\Psi
    \,,
\eea
which means that the operator $J_{z}$ does not commute with the squared of the Hamiltonian and in addition,
the vortical dipole term is absent.

%%%%%%%%%%%%%%%%%%%%%%%%%%%%%%%%%%%%%%%%%%%%%%%%
%%%%%%%%%%%%%%%%%%%%%%%%%%%%%%%%%%%%%%%%%%%%%%%%
%%%%%%%%%%%%%%%%%%%%%%%%%%%%%%%%%%%%%%%%%%%%%%%%
\subsection{Dirac Equation}\label{subsec2}
The expression for the Dirac equation in
a general curved background was already 
mentioned in Eq.\ (\ref{Dirac}).
To consider the general case, we also need to add a chemical potential to the equation. This requires
understanding of time-like vectors in our metric 
\be\label{metric}
ds^2=-dt^2+\Big(dx+\beta^{x}(y)dt\Big)^2+dy^2+dz^2\,,
\ee
with $\beta^{x}(y)$ some arbitrary function of $y$. Note that in the rest of our discussion, we will
consider a simpler setup than what we used in the last section which assumes $\beta^{x}(y)\neq0$
but $\beta^{y}(x)=0$. 
Different configurations of this sort just change the prefactor
of $\omega^{z}$ in the equations.  The  metric in Eq.\,(\ref{metric}) is a rectilinear model of the
Kerr metric and has interesting properties \cite{Fulling}; the hypersurfaces of constant $t$ are
spacelike and they satisfy Cauchy problem. In addition, the metric and the dynamics of fields are also
time independent. The stationary condition can be violated only if $|\beta^{x}(y)|\geq1$ for some $y$.
It's the result of the time independence of the problem that the norm of the normal vector to the equal-time
hypersurfaces, $\partial_{\mu}t=(1,0,0,0)$, is greater than zero and time is an ignorable coordinate. 
We keep in mind that, contravariant components of the Killing vector responsible for time translations
are $\partial_{t}x^{\mu}=(1,0,0,0)$ with $g_{00}=-1+\beta_{x}^{2}$. Therefore, when 
the condition $|\beta^{x}|>1$ is satisfied, the Killing vector is spacelike and the metric
is no longer stationary. But in this paper, we keep curved space corrections upto $\mathcal{O}(\beta_{x}^2)$
so our configuration is manifestly time-independent and stationary.

We seek a solution of $\Psi(t)=\chi e^{-i\varepsilon t}$ for $\varepsilon>0$. 
Adding chemical potential, $-i\mu\gamma^{0}$ to the equations of motion, it is understood that
\be
   \left(
    -i\varepsilon \gamma^{0}-i\mu\gamma^{0}
    +\half \left\{\gamma^{i},\frac{\partial}{\partial x^{i}}\right\}
    +\frac{1}{4}\left\{\gamma^{\mu},G_{[\alpha\beta]}\right\}\Sigma^{\alpha\beta}_{\quad\mu}\right)\chi=0\,.
\ee
Simplification of the above equation gives 
\bea\label{diag-dirac}
\left( 
    \begin{array}{cc}
    0 & 
     -i(\varepsilon\!+\!\mu)\!-\!\beta^{x}\partial_{x}\!+\!\sigma^{i}\partial_{x^{i}}\!-\!i\lambda\omega^{z}\sigma^{z} \\
     -i(\varepsilon\!+\!\mu)\!-\!\beta^{x}\partial_{x}\!-\!\sigma^{i}\partial_{x^{i}}\!-\!i\lambda\omega^{z}\sigma^{z}  & 
    0\\
    \end{array} \right)\chi\!=\!0\,,
\eea
 with $\lambda=\half$.
As we mentioned earlier, in general $\beta^{x}(y)$ is an arbitrary function of the $y$ coordinate. Requiring vorticity to 
be constant means that $\beta^{x}(y)$ has to be a linear function and in principle one has to solve the Dirac equation
in this external potential. However, we are interested at a regime
\footnote{
	We make this choice to neglect the effect of the
	superradiant modes i.e. reflected waves from a potential barrier.} 
that $\varepsilon\gg \beta^{x}(y)$. Then
normalizations of waves are not affected by this background and we can neglect the reflection of the waves from the potential. 
\begin{figure}[ht] \centering
\mbox{}
\epsfysize=2.4in
\leavevmode
\epsfbox{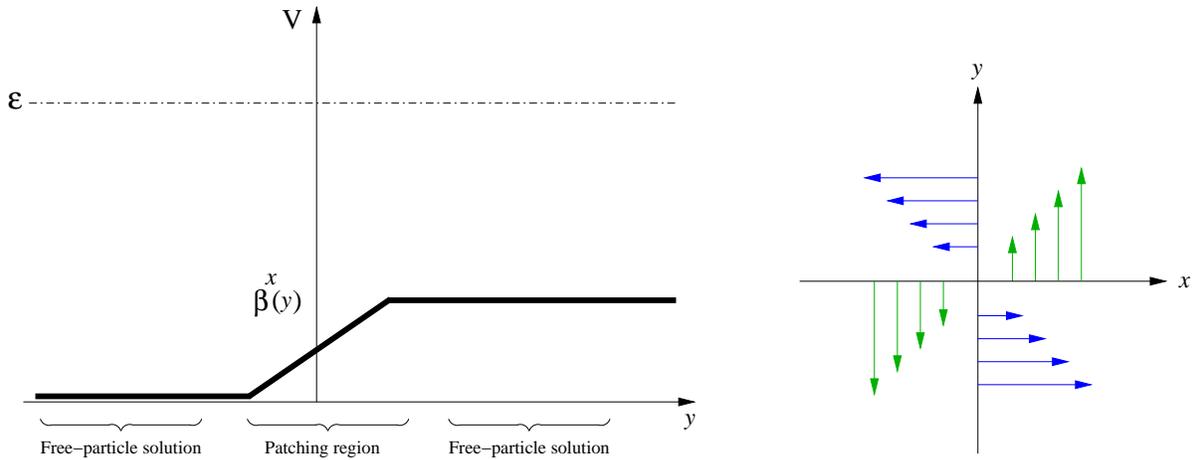}
\caption{(Left side) Dirac equation in the curved background is illustrated by spinors in a potential well. Throughout our
calculation we will be working in the semiclassical regime that WKB approximation is valid. (Right side) Picture of a rotating fluid
 made of  $\beta^{x}(y)$ and $\beta^{y}(x)$ profiles in blue and green colors respectively.}
\label{Fred}
\end{figure}

On both sides of the potential well, solutions of the Dirac equation are those of the free fermions
except that on the right-hand side,  energy eigenvalues are shifted by the constant value of
$\beta^{x}P_{x}$. Furthermore, Eq.\,(\ref{diag-dirac}) shows that transition from the left to 
the right region is accompanied by a discontinuity in the momentum $P^{z}$ by $\lambda\omega^{z}$.
Another way of saying this is that we use WKB approximation for the mid region.  Then the solutions of this problem are
simply obtained by the following replacements 
\bea
   &&u_{R}:\quad p^{0}\rightarrow p^{0}+\mu+\beta^{x}p_{x}\,,\quad p_{z}\rightarrow p_{z}-\lambda \omega^{z}\,,
\\
&&u_{L}:\quad p^{0}\rightarrow p^{0}+\mu+\beta^{x}p_{x}\,,\quad p_{z}\rightarrow p_{z}+\lambda \omega^{z}\,,
\eea
in the solutions for the right-handed and left handed free spinors. 

The fact that energy eigenvalues depend on $\beta^{x}$ bears more attention. One property of the metric
in Eq.\,(\ref{metric}) is that under the Galilean coordinate transformation of $x^{'}=x-\beta t$ with
$t^{'}=t$, $y^{'}=y$ and $z^{'}=z$, we will have a gauge transformation in the form
$\beta^{'}_{x}(y)\equiv\beta_{x}(y)+\beta$. This means that frequency and energy are determined up to
$\beta P^{x}$. As it has been pointed out by Fulling \cite{Fulling}, the energy difference between two states
with different $P^{x}$ is ill-defined, but the conservation of momentum in $x$ direction that is preserved 
by the Killing vector $\partial_{x}$, forbids such transitions. The energies
\bea\label{energies}
    \varepsilon_{R}+\mu+\beta^{x}p_{x}&=&\pm|\bm p|\mp\lambda\frac{p_{z}}{|\bm p|}\omega^{z}\,,
    \nonumber\\    
    \varepsilon_{L}+\mu+\beta^{x}p_{x}&=&\pm|\bm p|\pm\lambda\frac{p_{z}}{|\bm p|}\omega^{z}\,,
\eea
for the right-handed and left-handed spinors, are advantageously presented for the future calculations. 
There, upper and lower signs are for particles and antiparticles respectively. 

Since we are going to adopt the second-quantization approach  in the next section, we shall elaborate on the role played by the 
creation and annihilation operators. Essentially, they should be consistent with the above pattern for energies. For instance,
in terms of the expansion over momentum $\bm p$, operators $a^{\dagger}_{\bm p}$ and $a_{\bm p}$ create and annihilate particles
with the momenta  $p_{z}\pm\lambda \omega^{z}$ and the energies given in   Eq.\,(\ref{energies}).  To keep the discussion as simple
as possible we only consider the right-handed particles. 

%%%%%%%%%%%%%%%%%%%%%%%%%%%%%%%%%%%%%%%%%%%
%%%%%%%%%%%%%%%%%%%%%%%%%%%%%%%%%%%%%%%%%%%
%%%%%%%%%%%%%%%%%%%%%%%%%%%%%%%%%%%%%%%%%%%
\subsection{Modified Lorentz Generators}\label{subsec3}
In this section, we study the effect of vorticities on Lorentz generators. Since chemical  potentials
explicitly break the Lorentz invariance we assume that $\mu=0$ in the following section. 

 In an arbitrary background,
Lorentz generators of boosts and rotations,
do not satisfy their flat spacetime
commutation relations of $\text{SO(3,1)}$ globally,
\be
    \left[M_{\mu\nu},M_{\sigma,\tau}\right]=\eta_{\mu\sigma}M_{\nu\tau}
					  +\eta_{\nu\tau}M_{\mu\sigma}
					  -\eta_{\mu\tau}M_{\nu\sigma}
					  -\eta_{\nu\sigma}M_{\mu\tau}\,,
\ee
with 
\bea\label{j}
    M_{\alpha\beta}=-\int_{\Sigma} d\Sigma_{\rho}
    \mathcal{M}_{\alpha\beta}^{\quad\rho}\,,
\eea
here $\Sigma$ is any spacelike Cauchy hypersurface with
the differential element of $d\Sigma_{\rho}\equiv d\Sigma\, n_{\rho}$.
As explained before,  surfaces of constant $t$ are spacelike and the normal
vectors to the  equal-time surfaces are given by $n_{\rho}=\partial_{\rho}t=(1,0,0,0)$.
The generalized angular momentum density, $\mathcal{M}_{\alpha\beta}^{\quad\rho}$  is defined \cite{Itzykson} by 
\be\label{tot-angul}
     \mathcal{M}_{\alpha\beta}^{\quad\rho}=-\left(x_{\alpha}T^{\;\;\rho}_{\beta}-x_{\beta}T^{\;\;\rho}_{\alpha}\right)\,,
\ee
with
\be\label{T}
    T^{\rho\alpha}=-\frac{ig^{1/2}}{4}\Big(
    \nabla^{\rho}\bar{\Psi}\gamma^{\alpha}\Psi
    +\nabla^{\alpha}\bar{\Psi}\gamma^{\rho}\Psi
    -\bar{\Psi}\gamma^{\rho}\nabla^{\alpha}\Psi
    -\bar{\Psi}\gamma^{\alpha}\nabla^{\rho}\Psi\Big)\,,
\ee
here $T^{\mu\nu}$ is the stress tensor density defined by 
$T^{\mu\nu}=\frac{\delta S}{\delta e_{\alpha\mu}}e_{\alpha}^{\,\nu}$ with the action given in 
Eq.\,(\ref{Dirac-log}). This is the equivalent
definition of the energy-momentum tensor density $T^{\mu\nu}=2\frac{\delta S}{\delta g_{\mu\nu}}$ for 
fermionic fields in curved spacetime. In Eq.\,(\ref{T}),  covariant derivatives 
are defined in Eq.\,(\ref{nabla}). The terms in Eq.\,(\ref{tot-angul}) comprise
both the orbital angular momentum and the spin. 
In the flat spacetime\footnote{
		We restrict our discussion
		to $t=0$ for simplicity.
},
 these generators are known for free spinors ($p^{0}=|\bm p|$) 
\be\label{flat-boost} 
    M_{\alpha\beta}=\frac{i}{2}\int_{\bm p}a^{\dagger}_{\bm p}u^{\dagger}_{R}(\bm p)
    \Big[
    L_{\alpha\beta}(\bm p)
    -2G_{[\alpha\beta]}
    \Big]
    u_{R}(\bm p)a_{\bm p}\,,
\ee 
where here 
\be
    L_{\alpha\beta}(\bm p)=\overleftrightarrow{p_{\alpha}\frac{\partial}{\partial p^{\beta}}}
			      -\overleftrightarrow{p_{\beta}\frac{\partial}{\partial p^{\alpha}}}
    \,,
\ee
and by $\leftrightarrow$ on top of an operator, we mean the antisymmetric
action on the right and left side of the operator. 
The coefficients in the expansion are defined based on the creation and the
annihilation operators for the right-handed fermions. As we discussed
in the previous section, we consider solutions for spinors that are almost those of the flat background
except that in a mid transition region, energy eigenvalues are slowly varying functions of $\beta^{x}(y)$.
This will also give rise to a nonzero derivative of the potential $\omega^{z}\equiv-\half\partial_{y}\beta^{x}$. 

So how do Lorentz generators look like in this geometry? To answer this question 
we study the definitions mentioned in Eq.\,(\ref{tot-angul}) and Eq.\,(\ref{T}). Our new spinors are
similar to the ones in flatspace but
momentum arguments are now in terms of the shifted momenta 
$p_{z}-3\lambda\omega^{z}$.  Another contribution comes from the  spin connection
in  covariant derivatives of Eq.\,(\ref{T}).  We look into the generators of boosts and rotations
separately.

\emph{Boosts.}---As a first example, we will look into the boost
along the $z$ direction, $M^{0z}\!=\!\int d^3\bm x \left(t T^{0z}\!-\!zT^{00}\right)$.
Below we study each of these terms using the  second-quantized formulation.
Although we restrict
ourselves to $t=0$, it's instructive to start with $T^{0z}$. 
Its component from 
Eq.\,(\ref{T}) explicitly reads ($g^{1/2}=-i$),
\bea\label{T0z}
    T^{0z}=\frac{-1}{4}\bar{\Psi}\left[
	   \gamma^{z}\left(\overset{\leftrightarrow}{\partial_{0}}-\beta^{x}\overset{\leftrightarrow}{\partial_{x}}\right)   
	   -\gamma^{0}\overset{\leftrightarrow}{\partial_{z}}
	   +\left\{G_{xy},\gamma^{z}\right\}\Sigma^{xy}_{\hspace{3mm}0}\right]\Psi\,,
\eea 
here gamma matrices with the coordinate indices are equal to ones carrying frame indices and
we have rewritten the derivatives in terms of their covariant components.  Expansion
over right-handed spinors, neglecting antiparticles, gives 
\be\label{vac-rot2}
    \frac{1}{2}\int_{\bm p}u^{\dagger}_{R}(\bm p)a^{\dagger}_{\bm p}
    \left[
    \sigma^{z}(p^{0}+\beta^{x}p_{x})+p^{z} +\lambda \omega^{z} 
\right]u_{R}(\bm p)a_{\bm p}\,,
\ee
in which  $\int_{\bm p}\equiv\int\frac{d^3\bm p}{(2\pi)^3}$.  
To take the volume integral, we have assumed that variation of $\beta^{x}(y)$ is ignorable 
comparing to the oscillatory Fourier factor (WKB approximation). 
We can compute $u^{\dagger}_{R}(\bm p)\sigma^{z}u_{R}(\bm p)$ using the Gordon identity
\bea\label{gordon}
    \Big(|\bm p|+|\bm q|\Big)u_{R}^{\dagger}(\bm q)\sigma^{z}u_{R}(\bm p)=
	    u^{\dagger}_{R}(\bm q)
	    \left(
	    q^{z}+p^{z}
	    +\half[\sigma^{z},\sigma^{j}](p_{j}-q_{j})\right)u_{R}(\bm p)\,.
\eea
Similar to the above identity, the sandwich of $\sigma^{x}$ and $\sigma^{y}$ is evaluated by replacing 
$|\bm p|$ and $|\bm q|$ with $|\bm p|-\lambda\omega^{z}\hat{p}_{z}$ and 
$|\bm q|-\lambda\omega^{z}\hat{q}_{z}$ on the left-hand side accordingly and  
trivially replacing the $z$ indices with $x$ and $y$ on the right-hand side.  
Suitable simplifications of Eq.\,(\ref{vac-rot2}) enable us to write it in the suggestive form
\be
  \int d^3\bm x\; T^{0z} = \half\int_{\bm p}u^{\dagger}_{R}(\bm p)u_{R}(\bm p)
    \Big[
    p_{z}+\lambda \omega^{z} -\hat{p}_{z} p_{0} 
    \Big]a^{\dagger}_{\bm p}a_{\bm p}\,,
\ee
with $-p_{0}\equiv p^{0}+\beta^{x}p_{x}$.

Another non trivial contribution to the boost operator,
$M^{0z}$, comes from  the term $\int d^3\bm x\; zT^{00}$ with $T^{00}$ from Eq.\,(\ref{T}) since the spin connection
is not zero, $\Sigma^{xy}_{0}=-\omega^{z}$, as it's clear from Eq.\,(\ref{T0z}) by replacing 
$z$ with $0$ coordinate. After Fourier expansion and decomposing different modes, the integral 
reads\footnote{To avoid defining new notations, we will constantly use $G_{[\alpha\beta]}$ 
	       sandwiched by the right-handed spinors, in spite of the fact that they are doublets.}   
\bea
    \int d^3\bm x\; zT^{00}=
    \half\int d^3\bm x\; z\int_{\bm p, \bm q}\hspace{-3mm}
    a^{\dagger}_{\bm q}u_{R}^{\dagger}(\bm q)
    \Big[
    -p_{0}-q_{0}+\omega^{z}\left\{\gamma^{0},G_{[xy]}\right\}
    \Big]
    u_{R}(\bm p)a_{\bm p}
    e^{i\bm x\cdot(\bm p-\bm q)}\,.
\eea
Applying $\left\{\gamma^{0},G_{[xy]}\right\}=i\sigma^{z}\gamma^{0}$ and use of
the following identity,
\be
\int d^3\bm x\, x^{j}\int_{\bm p,\bm q}  e^{i\bm x\cdot(\bm p-\bm q)}f(\bm p,\bm q)=
\frac{i}{2}\int_{\bm p,\bm q}\delta^{3}(\bm p-\bm q)
\left(\frac{\partial}{\partial p_{j}}-\frac{\partial}{\partial q_{j}}\right)f(\bm p,\bm q)\,,
\ee
yields
\bea
    \int d^3\bm x\; zT^{00}=
    -\frac{i}{2}\int_{\bm p}
    \!\!
    a^{\dagger}_{\bm p}u_{R}^{\dagger}(\bm p)
    \left[
    \overleftrightarrow{p_{0}\frac{\partial}{\partial p_{z}}}
    +\frac{\omega^{z}\sigma^{z}}{4}\overleftrightarrow{\frac{\partial}{\partial p_{z}}}
    \right]
    u_{R}(\bm p)a_{\bm p}\,.
\eea
Our argument in this
calculation was general and we could generalize it to $M^{0x}$ and $M^{0y}$,
\bea\label{boost-notflat}
    M^{0i}=
   \frac{i}{2}\int_{\bm p}
    \!\!
    a^{\dagger}_{\bm p}u_{R}^{\dagger}(\bm p)
    \left[
    \overleftrightarrow{p_{0}\frac{\partial}{\partial p_{i}}}
    +\frac{\omega^{z}\sigma^{z}}{4}\overleftrightarrow{\frac{\partial}{\partial p_{i}}}
    \right]
    u_{R}(\bm p)a_{\bm p}\,,
\eea
and in the above, the first term has the form similar to $M_{\text{flat}}^{0i}$ in Eq.\,(\ref{flat-boost}) which is 
the contribution in the absence of the vorticities.

\emph{Rotations.}---It's also interesting to extend our computation to generators of rotations. 
The simplest case is $M^{xy}$, rotations around the vorticity. As we have seen in Sec.\,\ref{sec2}
the component of the angular momentum along the direction of vorticities commutes with the Hamiltonian.
Simple calculation here also confirms that this generator is not modified by vorticity corrections
$M^{xy}\equiv M^{xy}_{\text{flat}}$. But this is certainly not true for $M^{xz}$ and $M^{yz}$. 
Similar computation gives
\bea\label{rotation-notflat}
    M^{xz}=
	  M^{xz}_{\text{flat}}
	  +\frac{i\omega^{z}}{8}\int_{\bm p}
	  a^{\dagger}_{\bm p}u_{R}^{\dagger}(\bm p)
	  \left[\left(1-\hat{p}^{2}_{z}\right)\overleftrightarrow{\frac{\partial}{\partial p_{x}}}
	  +\frac{2\hat{p}_{z}}{|\bm p|}G_{[xz]}\right]
	  u_{R}(\bm p)a_{\bm p}\,,
\eea 
with  the identical expression\footnote{Also note that by $M^{xy}_{\text{flat}}$ or 
					$M^{xz}_{\text{flat}}$, we mean the their forms are identical to
					their corresponding flat-spacetime counterparts, otherwise their
					expansions are in terms of the creation and annihilation operators
					prescribed below Eq.\,(\ref{energies}).} for 
$M^{yz}$  by a simple interchange of $x\leftrightarrow y$. 

We would like to give emphasis to the role played by $T^{0z}$ in 
the structure of the boosts and rotations. Since this component of the stress tensor has an 
extra contribution proportional to the vorticity, commutation relations will be modified. 
We postpone further study of this issue to future works.

\subsection{Chiral Vortical Effect}\label{subsec4}
So far we have covered different aspects of the fermionic theories in the presence of the background vorticity.
One novel aspect in this regard is the chiral vortical effect.
The electric current density deduced from the variation of the action in Eq.\,(\ref{Dirac-log}),
\begin{equation}
    \mathcal{L}=-ig^{1/2}\bar{\Psi}\gamma^{\mu}
    \left(\frac{\partial}{\partial x^{\mu}}-ieA_{\mu}
    +\half G_{[\alpha\beta]}\Sigma^{\alpha\beta}_{\mu}\right)
    \Psi\,,
\end{equation}
with respect to the electromagnetic four-potential $A_{\mu}(x)$ is $j^{\mu}=-g^{1/2}e\bar{\Psi}\gamma^{\mu}\Psi$. Specifically
the $z$ component, the direction along the vorticity, of the total
current for the right-handed fermions reads
\bea
J^{z}&=&e\int d^3\bm x\int_{\bm p,\bm q}
	  \left[
	  u^{\dagger}_{R}(\bm q)\sigma^{z} u_{R}(\bm p)
	  a^{\dagger}_{\bm q}a_{\bm p}\right]
	  e^{i\bm x\cdot(\bm p-\bm q)}\,.
\eea
We can use the Gordon's identity in Eq.\,(\ref{gordon}) to simplify the above expression. 
Successive substitutions indicate that
\be\label{current}
    J^{z}=e\int d^3\bm x\int_{\bm p,\bm q}u^{\dagger}_{R}(\bm q)a^{\dagger}_{\bm q}
    \left[
    \frac{p^{z}+q^{z}}{|\bm p|+|\bm q|}
    -i\frac{p^{x}+q^{x}}{|\bm p|+|\bm q|}\frac{p^{y}-q^{y}}{|\bm p|+|\bm q|}
    \right]
    u_{R}(\bm p)a_{\bm p}
    e^{i\bm x\cdot(\bm p-\bm q)}\,,
\ee
with corrections of $\mathcal{O}(\Delta p^2)$ with $\Delta p\!\equiv\!p\!-\!q$ that are suppressed. 
To establish a connection with CVE, we compute the thermodynamic average by
\be
    \langle J^{z}\rangle=
    \frac{\sum_{\alpha}\exp\{-\frac{\varepsilon_{\alpha}}{T}\}\langle\alpha|J^{z}|\alpha\rangle}
    {\sum_{\alpha}\exp\{-\frac{\varepsilon_{\alpha}}{T}\}}\,,
\ee
where each state is marked with a label $\alpha$ and the corresponding energy is  $\varepsilon_{\alpha}$. 
In the absence of the vorticities, this gives
$\langle a^{\dagger}_{\bm q}a_{\bm p}\rangle=\delta^{3}({\bm q-\bm p})f(|\bm p|)$ with 
$f(\omega)\equiv\left(e^{\frac{\omega}{T}}-1\right)^{-1}$. But as intended  
in Section.\,\ref{sec2}, the influence of the vorticities  change the dispersion relation
and for the distribution function, we have
$f\left(|\bm p|-\beta^{x}(y)p_{x}-\lambda\omega^{z}\hat{p}_{z} \right)$.
Rewriting $p^{y}-q^{y}$ in the coordinate space and expanding the anisotropic distribution
function over its argument, we can express Eq.\,(\ref{current}) as
\be
    \langle J^{z}\rangle=e\int d^3\bm x\int_{\bm p,\bm q}
    u^{\dagger}_{R}(\bm q) u_{R}(\bm p)
    e^{i\bm x\cdot(\bm p-\bm q)}\left.\frac{\partial f}{\partial \varepsilon}\right|_{f=f_{0}}\delta(\bm p-\bm q)    
    \left[
    \frac{p^{z}+q^{z}}{|\bm p|+|\bm q|}
    -\frac{p^{x}+q^{x}}{\left(|\bm p|+|\bm q|\right)^2}\frac{\partial}{\partial y}
    \right]
   \varepsilon_{R}\,,
\ee
where $f_{0}$ displays the equilibrium distribution function. Since $\partial_{y}$ only acts on $\beta^{x}$
giving a factor of $\omega^{z}=-\frac{\partial_{y}\beta^{x}}{2}$ and hence a constant, the volume integral 
would be trivial. Integration over $\bm q$  then yields
\be
    \frac{\langle J^{z}\rangle}{V}=e\,\omega^{z}\int_{\bm p}
    u^{\dagger}_{R}(\bm p)u_{R}(\bm p)
    \left.\frac{\partial f}{\partial \varepsilon}\right|_{f=f_{0}}    
    \left(
    -\frac{\hat{p}_{z}^{2}}{2}
    -\hat{p}_{x}^{2}
    \right)\,,
\ee
with $\int d^3\bm x=V$. This is the expression for CVE obtained for a massless right-handed spin-$\half$ particle. A similar
statement applies to the left-handed spin-$\half$ particles, with an opposite overall sign.
A derivation based on rotating frames and phase-space action has appeared recently in \cite{Chen:2014cla}.

\section{Summary}

In the forgoing discussion,  we studied the chiral fermions under the influence of the vorticities.
Our analysis is intrigued by the recent advances in the theory of hydrodynamics and treatments used
in this work is no surprise for experts on the Kerr black hole. Using the standard methods
of quantum field theory in curved spacetime, we derive
the Lagrangian, Hamiltonian and the equations of motion for the chiral fermions.
Tentatively, we constructed $H^2$, the squared of the Hamiltonian. This was primarily important
since similar operation  in the flat spacetime gives the Hamiltonian, corresponding to the 
Schr\"{o}dinger equation for fermions, which is devoid of the spin degrees of freedom.
In this case, endowed with the  vorticities,  a similar interaction to the medium under a background
magnetic field appears. 
The chemical potential times the Pauli sigma matrices acts as a dipole moment under the vorticities.
It is also interesting to observe that having two
vorticities in opposite directions do not totally degrade or neutralize each other. Some specific 
interactions that appear in $H^{2}$ are due to moving frames and independent of the former directions.

Rather than looking for less soluble models, we have pursued the Dirac  equation for energies
$\varepsilon\gg\beta^{x}(y)$. This allows us to deploy the WKB approximation  to solve for the 
solutions in this background that brought a firmer physical interpretation. 
For instance, for the right-handed spinors, substitution of 
$p^{0}\rightarrow p^{0}+\mu+\beta^{x}p_{x}$ and $p_{z}\rightarrow p_{z}-\lambda \omega^{z}$, with 
$\lambda=\half$ manifests the  solution for the right-handed spinors under the action
of the vorticities. 
The new dispersion relation for $\varepsilon_{R}$ agrees with \cite{Chen:2014cla}
while those authors reasoned based on rotating frames.  We point out again that as far as we are in 
the regime of the validity of the WKB approximation, $\beta^{x}(y)$ can have any arbitrary shape.

Another invoking argument was based on the generators of boosts and rotations under the ambient vorticities.
Our computation is a prototype of its kind in this context. Here there has been a fundamental assumption
about the configurations that we pick up. We only consider the static and stationary backgrounds.
Hence, we can use the second quantization.  Since the above mentioned operators are defined based on
stress tensors, one is curious about the role that $T^{0z}$ plays while having $\omega^{z}\neq0$.
Direct calculation shows that there are two contributions coming from this term. One is due to the fact 
that $p^{z}$ has a shift proportional to the vorticity and in addition the spin connection
$\Sigma^{xy}_{\quad0}$ is not zero. The 
final conclusion will be that all generators comprised of $T^{0z}$, are subject to  corrections
proportional to the vorticity to first order. 

A fascinating continuing body of work has been invested on the topic of the chiral vortical effects in recent years.
Our argument is intimately tied to the rudimentary form of the Lagrangian. One important step is 
to pay attention to the creation and annihilation operators in the background
of the vorticities, since these operators now create and annihilate particles out of the vacuum with shifted 
energies and momenta. This is signified in  the average ensemble and the distribution function that 
thereby  will be a function of $f\left(|\bm p|-\beta^{x}(y)p_{x}-\lambda\omega^{z}\hat{p}_{z} \right)$. 
The first term 
is of course the energy of the massless spinors in the  flat spacetime. The second and third terms, $\bm u\cdot\bm p$ and
$\lambda\bm\omega\cdot\bm \hat{p}$, are corrections to the energy of an observer in a rotating frame.   

In short, it may well be that  similar to the case of the electromagnetism that implementation
of the electromagnetic four-potentials  admits one to study such  systems, methods in curved spacetime are the
proper way of probing relativistic field theories under the influence of the vorticities.  

To close the discussion, we point out again that in this paper small and constant vorticities are considered. 
Extensions to more general cases are accompanied by the inclusion of superradiant and Klein modes. It is also very
important to make a connection to non-equilibrium physics such as turbulence \cite{Adams}, for
a better understanding of the holographic picture.  

\begin{acknowledgements}
I would like to thank Dam T. Son and Mikhail A. Stephanov for discussion when I began this work. 
Discussions with Matthias Blau and Adolfo Guarino were also of help during the course of this work.
I would also like to thank Mikko Laine and Guy D. Moore for their support. 
This work was partly supported by the Swiss National Science Foundation
(SNF) under grant 200021-140234. 
\end{acknowledgements}

\end{document}